\documentclass[11pt]{article}
\usepackage{epsf} 
\usepackage{amsmath}
\usepackage{amssymb}
\usepackage{epsfig}
\usepackage{latexsym}
\usepackage{amsfonts}
\usepackage{graphicx}%
\usepackage{varioref}
\usepackage{ifthen}
\begin{document}
\parindent 0mm 
\setlength{\parskip}{\baselineskip} 
\thispagestyle{empty}
\pagenumbering{arabic} 
\setcounter{page}{0}
\mbox{ }
\rightline{UCT-TP-280/2010}
\newline
\newline
\rightline{May 2010}
\newline

\begin{center}
\Large \textbf{Determination of light quark masses in QCD }
{\LARGE \footnote{{\LARGE {\footnotesize Plenary talk at the Conference in honour of Murray Gell-Mann's 80th birthday, Nanyang Technical University, Singapore, February 2010. To be published in the proceedings (Mod. Phys. Lett. A). This talk is based on work done in collaboration with N. F. Nasrallah, R. H. R\"{o}ntsch, and K. Schilcher.
Work supported in part by NRF (South Africa). }}}}
\end{center}
\begin{center}
C. A. Dominguez
\end{center}

\begin{center}
Centre for Theoretical Physics and Astrophysics\\
University of Cape Town, Rondebosch 7700, South Africa, and Department of Physics, Stellenbosch University, Stellenbosch 7600, South Africa
\end{center}

\begin{center}
\textbf{Abstract}
\end{center}

\noindent
The standard procedure to determine (analytically) the values of the quark masses is to relate QCD two-point functions to experimental data in the framework of QCD sum rules. In the case of the light quark sector, the ideal Green function is the pseudoscalar correlator which involves the quark masses as an overall multiplicative factor. For the past thirty years this method has been  affected by systematic uncertainties originating in the hadronic resonance sector, thus limiting the accuracy of the results. Recently, a major breakthrough has been made allowing for a considerable reduction of these systematic uncertainties and leading to light quark masses accurate to better than 8\%. This procedure will be described in this talk  for the up-, down-, strange-quark masses, after a general introduction to the method of QCD sum rules. 

\newpage

\section{Introduction}
Due to quark and gluon confinement in QCD the (analytical) determination of the values of the quark masses necessarily calls for an approach different from that for ordinary, non confined particles. The ideal approach is to  consider a QCD correlation function which on the one hand involves the quark masses and other QCD parameters, and on the other hand it involves a measurable (hadronic) spectral function. Given  a framework to relate both representations of this correlator, i.e. the QCD and the hadronic representation, the quark masses would then become a function of QCD parameters, e.g. the strong coupling, some vacuum condensates reflecting confinement, etc., and measurable hadronic parameters. This framework has traditionally been that of the QCD sum rules \cite{REVIEW}, to be described in the following sections.
Unfortunately, the directly measurable Green functions in the light quark sector are the vector and axial-vector correlators entering $\tau$-decays, and  which involve the quark masses as sub-leading contributions. This feature impacts negatively on the accuracy of the results. The pseudoscalar correlator, which contains the quark masses as an overall multiplicative factor, thus making it the ideal object to be used, is not realistically measurable beyond the ground state pseudoscalar meson pole (pion or kaon). In the case of the scalar correlator the situation is worsened by the absence of a ground state pole. The standard approach for close to thirty years in connection with the (pseudo) scalar correlator has been to model the hadronic resonance spectral functions as best as possible. In the early days, radial excitations of the ground state pseudoscalar meson were parametrized in zero-width with model dependent couplings, followed later by finite-width parameterizations,  and eventually incorporating sound threshold constraints from chiral perturbation theory \cite{CAD0}-\cite{PRADES}. At present, the masses and widths of the first two radial excitations of the pion and the kaon are well known experimentally. However, the model dependency is still unavoidable as inelasticity, non-resonant background and resonance interference are realistically impossible to guess.\\
As a result of this model dependency of the hadronic sector, light quark mass determinations have been historically affected by systematic uncertainties. A major breakthrough has been achieved recently \cite{DNRS1}-\cite{DNRS2} by using a successful procedure which reduces considerably the contribution of the unknown resonance sector. This quenching is accompanied by a corresponding enhancement of the better known terms in the correlator, i.e. perturbative QCD and the pseudoscalar meson pole. As a result, the light quark masses can now be determined (analytically) with an unprecedented accuracy of better than 8\%, as will be described in this talk.

\section{QCD Sum Rules}
The method of QCD sum rules, introduced by Shifman, Vainshtein, and Zakharov \cite{SVZ} more than thirty years ago, has become a powerful technique to study hadronic physics in the low energy resonance region by means of QCD \cite{REVIEW}. This method is also a complementary tool to numerical simulations of QCD on a lattice \cite{LATTICE}. The range of applications has steadily grown over the years and covers the determination of the hadronic spectrum (masses, couplings, and widths), as well as electromagnetic, weak, and strong form factors. QCD sum rules have also been extended to finite temperature and density \cite{T1}, thus allowing for a study of chiral symmetry restoration and quark-gluon deconfinement \cite{T2}. On the QCD sector, sum rules are employed to extract the values of the quark masses \cite{REVIEW} and of the strong coupling at a scale of the $\tau$-lepton mass \cite{ALEPH}-\cite{PICH}, and they are also a tool to confront QCD predictions with experimental data, e.g. from $e^+ e^-$ annihilation and hadronic $\tau$-lepton decays. This method is based on two fundamental pillars: (i) the operator product expansion of current correlators at short distances, extended beyond perturbation theory to incorporate quark-gluon confinement, and (ii) Cauchy's theorem in the complex energy (squared) plane, often referred to as quark-hadron duality. To be more specific, let us consider a typical object in QCD in the form of the two-point function, or current correlator
\begin{equation}
\Pi(q^2)\,=\,i\; \int \,d^4 x \; e^{iqx} \; <0|\,T(J(x)\,J(0))\,|0 >,
\end{equation}
where the local current $J(x)$ is built from the quark and gluon fields entering the QCD Lagrangian, and it has definite quantum numbers. Alternatively, this current can be written in terms of hadronic fields having the same quantum numbers. The specific  choice of current will depend on the application one has in mind. For instance, if one is interested in determining the hadronic properties of the $\rho^+$-meson, then one would choose the QCD vector isovector current $J_\mu(x) = \bar{d}(x) \, \gamma_\mu \,u(x)$, and its hadronic realization in terms of the $\rho^+$-meson field. If the goal is to determine the values of the light quark masses, then the ideal object would be the correlator involving the axial-vector current divergences $J_5(x)|^i_j = (m_i + m_j) \bar{\psi}^i(x)  \gamma_5 \psi_j(x)$, with $i,j$ the up, down, or strange quark flavors. The hadronic representation of this correlator contains the pseudoscalar meson ($\pi$ or K) mass and coupling, followed by its radial excitations and the hadronic continuum. The tool to relate  these two representations is Cauchy's theorem in the complex energy (squared) plane, to be discussed shortly. \\
The QCD correlator, Eq.(1), will contain a perturbative piece (PQCD), computed up to a given loop order in perturbation theory, and a non perturbative part mostly reflecting quark-gluon confinement. The leading order in PQCD is shown in Fig.1.  Since QCD has never been solved analytically, the effects due to confinement can only be introduced by parameterizing quark and gluon propagator corrections effectively in terms of vacuum condensates. This is done as follows. In the case of the quark propagator
\begin{equation}
S_F (p) = \frac{i}{\not{p} - m}\;\;\Longrightarrow \;\;\frac{i}{\not{p} - m + \Sigma(p^2)} \;, 
\end{equation}
the quark propagator correction $\Sigma(p^2)$ would contain the information on confinement. One expects this correction to peak at and near the quark mass-shell, i.e. for $p \simeq 0$ in the case of light quarks. Effectively, this can be viewed as in Fig. 2, where the (infrared) quarks in the loop have zero momentum and interact strongly with the physical QCD vacuum. This effect is then parameterized in terms of the quark condensate $\langle 0| \bar{q}(0) q(0) | 0 \rangle$.
\begin{figure}[ht]
\begin{center}
  \includegraphics[height=.1\textheight]{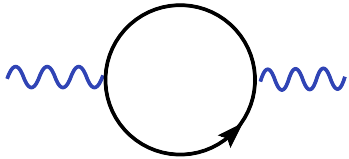}
  \caption{Leading order PQCD correlator. All values of the four-momentum of the quark in the loop are allowed. The blue wiggly line represents the current of momentum $q$.}
  \label{fig:figure1}
  \end{center}
\end{figure}
Similarly, in the case of the gluon propagator one would have
\begin{equation}
D_F (k) = \frac{i}{k^2}\;\;\Longrightarrow \;\;\frac{i}{k^2 + \Lambda(k^2)} \;,
\end{equation}
where the gluon propagator correction will peak at $k\simeq 0$, and the effect of confinement in this case will be parameterized by the gluon condensate $\langle 0| \alpha_s\; \vec{G}^{\mu\nu} \,\cdot\, \vec{G}_{\mu\nu}|0\rangle$ (see Fig.3).
\begin{figure}[ht]
\begin{center}
\includegraphics[scale=0.8]{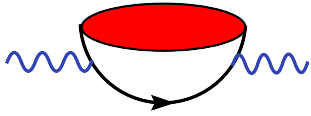}
\caption{Quark propagator modification due to (infrared) quarks interacting with the physical QCD vacuum, and involving the quark condensate. Large momentum flows only through the bottom propagator.}
\label{fig:figure2}
\end{center}
\end{figure}
In addition to the quark and the gluon condensate there is a plethora of higher order condensates entering the OPE of the current correlator at short distances, i.e.
\begin{equation}
\Pi(q^2)|_{QCD}\,=\, C_0\,\hat{I} \,+\,\sum_{N=0}\;C_{2N+2}(q^2,\mu^2)\;\langle0|\hat{O}_{2N+2}(\mu^2)|0\rangle \;,
\end{equation}
where $\mu^2$ is the renormalization scale, and where the Wilson coefficients in this expansion depend on the Lorentz indexes and quantum numbers of $J(x)$ and  of the local gauge invariant operators $\hat{O}_N$ built from the quark and gluon fields. These operators are ordered by increasing dimensionality and the Wilson coefficients, calculable in PQCD, fall off by corresponding powers of $-q^2$. 
Since there are no gauge invariant operators of dimension $d=2$ involving the quark and gluon fields in QCD, it is normally assumed that the OPE starts at dimension $d=4$ (with the quark condensate being multiplied by the quark mass). This is supported by results from QCD sum rule analyses of $\tau$-lepton decay data, which show no evidence of $d=2$ operators \cite{C2a}-\cite{C2c}.
\begin{figure}[ht]
\begin{center}
\includegraphics[scale=0.6]{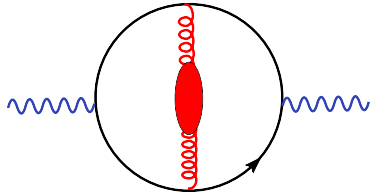}
\caption{Gluon propagator modification due to (infrared) gluons interacting with the physical QCD vacuum, and involving the gluon condensate. Large momentum flows only through the quark propagators.}
\label{fig:figure3}
\end{center}
\end{figure}
The unit operator in Eq.(4) has dimension $d=0$ and $C_0 \hat{I}$ stands for the purely perturbative contribution. The Wilson coefficients as well as the vacuum condensates depend on the renormalization scale. In the case of the leading $d=4$ terms in Eq.(4) the $\mu^2$ dependence of the quark mass cancels the corresponding dependence of the quark condensate, so that this contribution is a renormalization group (RG) invariant. Similarly, the gluon condensate is also a RG invariant quantity, hence once determined in some channel these condensates can be used throughout. At dimension $d=6$ there appears the four-quark condensate, obtained from Fig.1 at the next to leading order (one gluon exchange) and allowing all four quark lines to interact with the physical vacuum (see Fig.4). While this condensate has a residual renormalization scale dependence, this is so small that in practice it can be ignored. The four-quark condensate, while relatively small, is crucial to explain the large $\rho(770)$ - $a_1(1260)$ mass splitting. In most applications $- q^2$ is chosen large enough so that the condensates of higher dimension ($d \geq 8$) can be safely ignored.
\begin{figure}[ht]
\begin{center}
\includegraphics[scale=0.6]{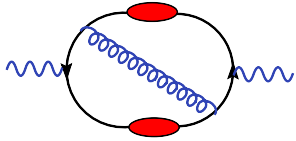}
\caption{The four-quark condensate of dimension $d=6$ in the OPE. This is responsible for the $\rho - a_1$ mass splitting. Large momentum flows only through the gluon propagator.}
\label{fig:figure4}
\end{center}
\end{figure}
The numerical values of the vacuum condensates cannot be calculated analytically from first principles as this would be tantamount to solving QCD exactly.
One exception is that of the quark condensate which enters in the Gell-Mann-Oakes-Renner relation, a QCD low energy theorem following from the global chiral symmetry of the QCD Lagrangian. Otherwise, it is possible to extract values for the leading vacuum condensates using QCD sum rules together with experimental data, e.g. $e^+ e^-$ annihilation into hadrons, and hadronic decays of the $\tau$-lepton. Alternatively, as lattice QCD  improves in accuracy it should become a valuable source of information on these condensates.\\
Turning to the hadronic sector, bound states and resonances appear in the complex energy (squared) plane (s-plane) as poles on the real axis, and singularities in the second Riemann sheet. In addition there will be multiple cuts reflecting non-resonant multi-particle production. All these singularities lead to a discontinuity across the positive real  axis. Choosing an integration contour as shown in Fig.5, and given that there are no further singularities in the complex s-plane, Cauchy's theorem leads to the finite energy sum rule (FESR)
\begin{figure}[ht]
\begin{center}
  \includegraphics[height=.25\textheight]{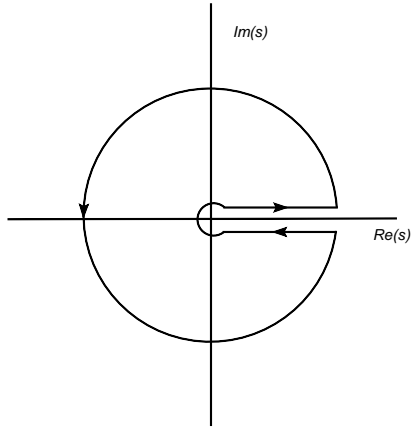}
  \caption{Integration contour in the complex s-plane. The discontinuity across the real axis brings in the hadronic spectral function, while integration around the circle involves the QCD correlator.}
\label{fig:figure5}
\end{center}
\end{figure}
\begin{equation}
\int_{\mathrm{sth}}^{s_0} ds\; \frac{1}{\pi}\; f(s) \;Im \Pi(s)|_{HAD} \; = \; -\, \frac{1}{2 \pi i} \; \oint_{C(|s_0|) }\, ds \;f(s) \;\Pi(s)|_{QCD} \;,
\end{equation}
where $f(s)$ is an arbitrary (analytic) function, $s_{th}$ is the hadronic threshold, and the finite radius of the circle, $s_0$, is large enough for QCD and the OPE to be used on the circle. Physical observables determined from FESR should not depend on $s_0$. In practice, though, 
this independence is not exact, and there is usually a region of stability where
observables are fairly independent of $s_0$, typically in the range $s_0 \simeq 1 - 3 \; \mbox{GeV}^2$. The variation of an observable in the stability region is incorporated into the error of the determination.
Equation (5) is the mathematical statement of what is usually referred to as quark-hadron duality. Since QCD is not valid in the time-like region ($s \geq 0$), in principle there is a possibility of problems on the circle near the real axis (duality violations). I shall come back to this issue later. The right hand side  of this FESR involves the QCD correlator which is expressed in terms of the OPE as in Eq.(4). The left hand side calls for the hadronic spectral function which is written as
\begin{equation} 
Im \Pi(s)|_{HAD}\,=\, Im \Pi(s)|_{POLE}\,+\, Im \Pi(s)|_{RES} \,\,+\, Im \Pi(s)|_{PQCD}\,\theta(s-s_0) \;,
\end{equation}
where the ground state pole, absent in some channels, is followed by the resonances which merge smoothly into the hadronic continuum above some threshold $s_0$. This continuum is expected to be well represented by PQCD if $s_0$ is large enough. Due to this, if one were to consider an integration contour in Eq.(5) extending to infinity, the cancellation between the hadronic continuum in the left hand side and the PQCD contribution to the right hand side, would render the sum rule a FESR. The performance of the contour integral in the complex s-plane is discussed in the next section.
\section{Finite energy sum rules}
The integration in the complex s-plane of the QCD correlator is usually carried out in two different ways, Fixed Order Perturbation Theory (FOPT) and Contour Improved Perturbation Theory (CIPT) \cite{CIPT1}-\cite{CIPT2}. The first method treats running quark masses and the strong coupling as fixed at a given value of $s_0$. After integrating all logarithmic terms ($\ln(-s/\mu^2)$) the RG improvement is achieved by setting the renormalization scale to $\mu^2 = - s_0$. In CIPT the RG improvement is performed before integration, thus eliminating logarithmic terms, and the running quark masses and strong coupling are integrated (numerically) around the circle. This requires solving numerically the RGE for the quark masses and the coupling at each point on the circle.
The FESR Eq.(5), with $f(s)=1$,  in FOPT can be written as
\begin{equation}
(-)^N \, C_{2N+2} \, \langle0| \hat{O}_{2N+2}|0 \rangle =  \int_0^{s_0} \,ds\, s^N \, \frac{1}{\pi}\, Im \,\Pi(s)|_{HAD} \,-\, s_0^{N+1} \; M_{2N+2}(s_0) \, ,
\end{equation}
where the dimensionless PQCD moments $M_{2N+2}(s_0)$ are given by
\begin{equation}
M_{2N+2}(s_0) = \frac{1}{s_0^{(N+1)}} \, \int_0^{s_0}\, ds\,s^N \, \frac{1}{\pi} \, Im \, \Pi(s)|_{PQCD}\;,
\end{equation} 
and $Im \,\Pi(s)$ is assumed dimensionless for simplicity.
\begin{figure}
[ht]
\begin{center}
\includegraphics[height=3.5 in, width=5.5 in]
{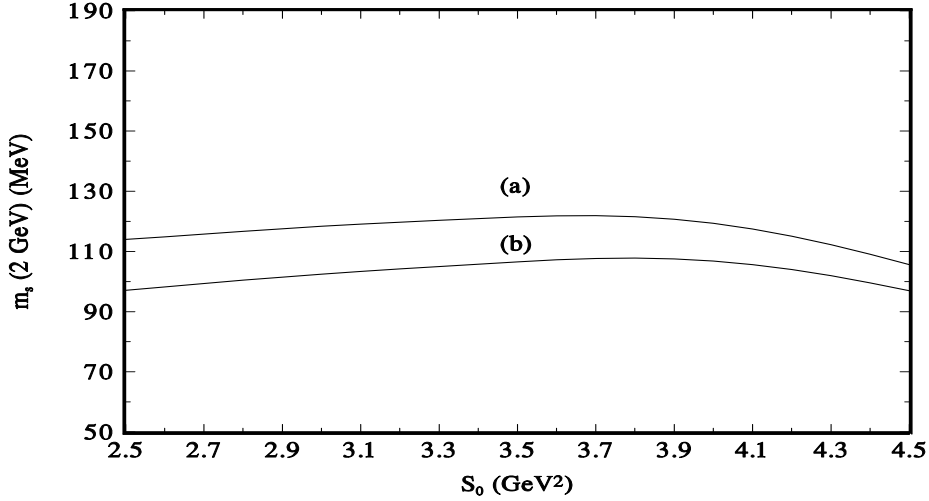}
\caption{Running strange quark mass in FOPT at a scale of 2 GeV as a function of $s_0$. Curve (a) is for $\Lambda = 330$ MeV, and curve (b) for $\Lambda = 420$ MeV, corresponding, respectively, to $\alpha_s(M_\tau^2) = 0.31$ and $\alpha_s(M_\tau^2) = 0.36$.}
\end{center}
\end{figure}

If the hadronic spectral function is known in some channel from experiment, then $Im \Pi(s)|_{HAD} \equiv Im \Pi(s)|_{DATA}$, and Eq.(7) can be used to determine the values of the vacuum condensates. Subsequently, Eq.(7) can be used in a different channel to determine the masses and couplings of the hadrons in that channel. It is important to mention that the correlator $\Pi(q^2)$ is generally not a physical observable. However, this has no effect in FOPT as the unphysical constants in the correlator do not contribute to the integrals. The situation is quite different in CIPT where Eq.(5) cannot be used for unphysical correlators. For instance for a correlator whose physical counterpart is the second derivative (needed to eliminate a first degree polynomial),  Cauchy's theorem and the resulting FESR must be written for the second derivative. In this case  one has to use the following identity \cite{DNRS1}-\cite{DNRS2}
\begin{equation}
\oint_{C(|s_0|) }\, ds \, g(s) \, \Pi(s) = \oint_{C(|s_0|) } \, ds \, [F(s) - F(s_0)] \;\Pi''(s) \;,
\end{equation}
where
\begin{equation}
F(s) = \int ^s ds' \left[ \int^{s'} ds'' g(s'') - \int^{s_0} ds'' g(s'')\right] \;,
\end{equation}
and $g(s)$ is an arbitrary function. This is easily proved by integrating by parts the right hand side of Eq.(9) and using Eq.(10) to obtain the left hand side. In this case Eq.(5) becomes
\begin{equation}
\int_{\mathrm{sth}}^{s_0} ds\; g(s)\;\frac{1}{\pi}\; Im \Pi(s)|_{HAD} \; = \; -\, \frac{1}{2 \pi i} \; \oint_{C(|s_0|) }\, ds \; [F(s)-F(s_0)] \;\Pi''(s)|_{QCD} \;,
\end{equation}
which is the master FESR to use in CIPT. The running quark masses and the running strong coupling entering $\Pi''(s)$  are now functions of the integration variable and are not fixed as previously in FOPT. 
The running coupling obeys the RGE
\begin{equation}
s \; \frac{d \, a_s(-s)}{d s} = \beta (a_s) = - \sum_{N=0} \beta_N \; a_s(-s)^{N+2} \;,
\end{equation}
where $a_s \equiv \alpha_s/\pi$, and  e.g. for three quark flavors $\beta_0 = 9/4$, $\beta_1 = 4$, $\beta_2 = 3863/384$,
$\beta_3 = (421797/54 + 3560 \zeta(3))/256$. In the complex s-plane $s = s_0\, e^{ix}$ with the angle $x$ defined in the interval $x \in (- \pi, \pi)$. The RGE then becomes
\begin{equation}
\frac{d \, a_s(x)}{d x} = - i \sum_{N=0} \beta_N \; a_s(x)^{N+2} \;,
\end{equation}
This RGE can be solved numerically at each point on the integration contour of Eq.(11) using e.g. a modified Euler method, providing as input $ a_s (x=0) = a_s (- s_0)$. Next, the RGE for the quark mass is given by
\begin{equation}
\frac{s}{m} \; \frac{d \, m(-s)}{d s} = \gamma (a_s) = - \sum_{M=0} \gamma_M \; a_s^{M+1} \;,
\end{equation}
where e.g. for three quark flavors $\gamma_0 = 1$, $\gamma_1 = 182/48$, $\gamma_2 = [8885/9 - 160 \,\zeta(3)]/64$, $\gamma_3 = [2977517/162 - 148720 \,\zeta(3)/27 + 2160 \,\zeta(4) - 8000\, \zeta(5)/3]/256$. With the aid of Eqs. (12)-(13) the above equation can be converted into a differential equation for $m(x)$ and integrated, with the result
\begin{equation}
m(x) = m(0) \;exp \Big\{ - i \int_0^x dx' \sum_{M=0} \gamma_M \, [a_s(x')]^{M+1}\Big\}\;,
\end{equation}
where the integration constant $m(0)$ can be  identified with $m(s_0)$.
\begin{figure}
[ht]
\begin{center}
\includegraphics[height=3.5 in, width=5.5 in]
{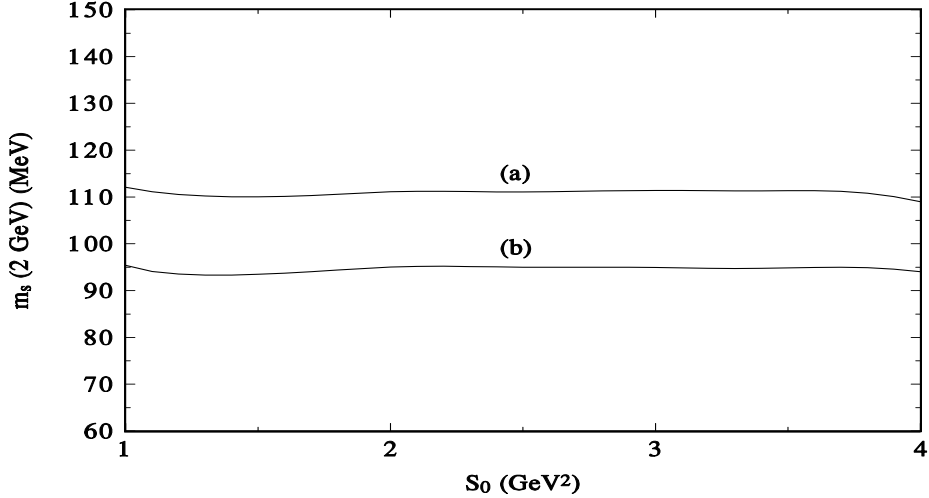}
\caption{Running strange quark mass in CIPT at a scale of 2 GeV as a function of $s_0$. Curve (a) is for  $\Lambda = 330$ MeV, and curve (b) for  $\Lambda = 420$ MeV, corresponding, respectively, to $\alpha_s(M_\tau^2) = 0.31$ and $\alpha_s(M_\tau^2) = 0.36$.}
\end{center}
\end{figure}

\section{THE LIGHT PSEUDOSCALAR CORRELATOR}
I discuss now the use of FESR to determine the values of the QCD light quark masses \cite{DNRS1}-\cite{DNRS2}. As mentioned in the Introduction, in this case  the ideal current operator in Eq.(1) is the axial-vector current divergence.
\begin{equation}
\psi_{5} (q^{2})   = i \, \int\; d^{4}  x \; e^{i q x} \; 
<|T(\partial^\mu A_{\mu}(x)|^j_i \;, \; \partial^\nu A_{\nu}^{\dagger}(0)|^j_i )|> \;,
\end{equation}
where $\partial^\mu A_{\mu}(x)|^j_i = (m_i + m_j) :\overline{q_j}(x) \,i \, \gamma_{5}\, q_i(x):\;$ is the divergence of the  axial-vector current, and $(i,j)$ are flavour indexes . To simplify the notation  we shall use in the sequel $m_i + m_j \equiv m$. The advantage is that the masses appear here as overall multiplicative factors, rather than as sub-leading power corrections like in other correlators, e.g. the vector or axial-vector correlators. The great disadvantage is that there is no direct experimental data beyond the pseudoscalar meson poles, i.e. the hadronic resonance spectral function, $Im \,\Pi(s)|_{RES}$ in Eq.(6) is not known experimentally. The only available information  is that there are two radial excitations in the non-strange ($\pi$) as well as in the strange ($K$) channel with known masses and widths. This is hardly enough to reconstruct the full spectral function. In fact, inelasticity, non-resonant background, and resonance interference are impossible to guess so that a model is needed for the resonant spectral function. This fact, which introduces a serious systematic uncertainty, has affected all quark mass determinations using QCD sum rules until recently \cite{DNRS1}-\cite{DNRS2}. The breakthrough has been to introduce an integration kernel in the FESR tuned to suppress substantially the resonance energy region above the ground state. This kernel is of the form 
\begin{equation}
 \Delta_5(s) = 1\, - \,a_0\,s \,-\,a_1\,s^2 \;, 
\end{equation}
where $\Delta_5(s)$ stands for either $f(s)$ in Eq.(5) for FOPT, or $g(s)$ in Eq.(9) for CIPT. The coefficients are fixed by requiring that $\Delta_5(s)$ vanish at the peak of the two radial excitations, i.e. $\Delta_5(M_1^2) = \Delta_5(M_2^2) = 0$. This has the effect of reducing the resonance contribution to the FESR to a couple of a percent of the ground state contribution, well below the uncertainty due to $\alpha_s$. Clearly, other more elaborate choices of the integration kernel are possible. It has been found, though, that the simplest form above is optimal in the sense of simplicity and of achieving the goal of reducing considerably the systematic uncertainty from the hadronic resonance sector.
\begin{figure}
[ht]
\begin{center}
\includegraphics[height=3.5 in, width=5.5 in]
{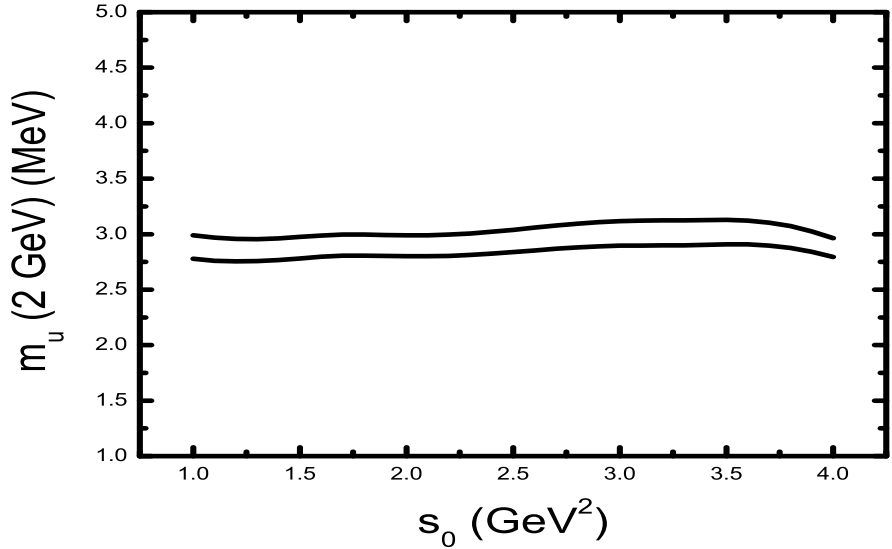}
\caption{Up quark mass at 2 GeV as a function of $s_0$ for $\alpha_s(M_\tau^2)= 0.335 (0.353)$, or
$\Lambda_{QCD} = 365 \;(397)\; \mbox{MeV}$, upper (lower) curve, respectively.}
\end{center}
\end{figure}

\begin{figure}
[ht]
\begin{center}
\includegraphics[height=3.5 in, width=5.5 in]
{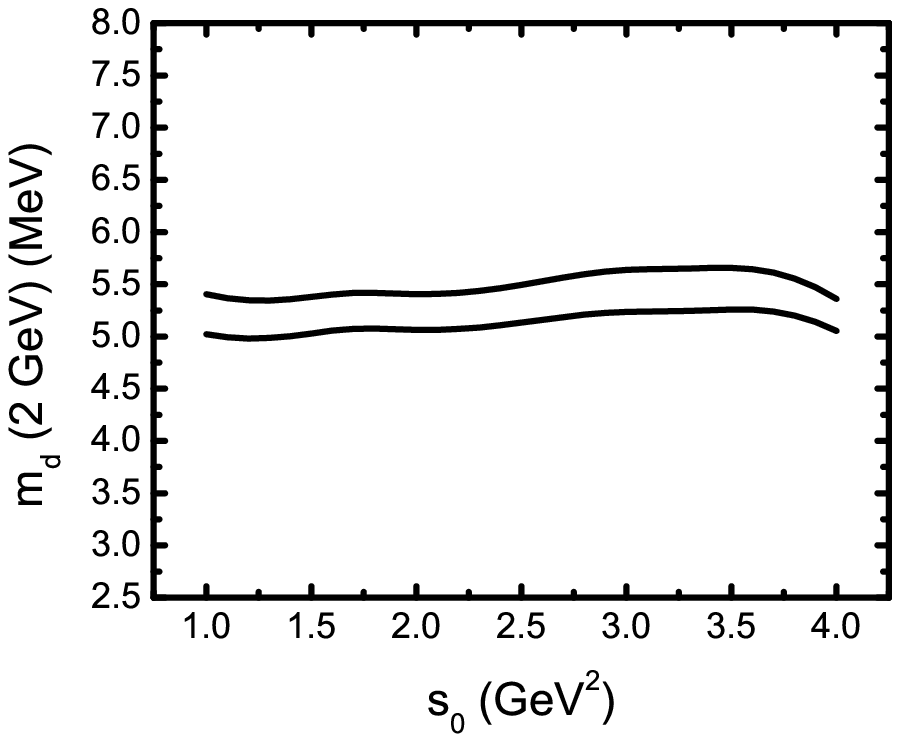}
\caption{Down quark mass at 2 GeV as a function of $s_0$ for  $\alpha_s(M_\tau^2)= 0.335 (0.353)$, or $\Lambda_{QCD} = 365 \;(397)\; \mbox{MeV}$, upper (lower) curve, respectively.}
\end{center}
\end{figure}

The FESR following from Cauchy's theorem, Eq.(5), takes now the form
\begin{eqnarray}
- \frac{1}{2\pi i}
\oint_{C(|s_0|)}
ds \;\psi_{5}^{QCD}(s)\; \Delta_5(s) &=& 2\; f_P^2 \; M_P^4\; \Delta_5(M_P^2)  \nonumber \\ [.3cm]
&+&
\int_{s_{th}}^{s_0}
ds \;\frac{1}{\pi} \;Im \;\psi_{5}(s)|_{RES}\;\Delta_5(s) \, , 
\end{eqnarray}
where $f_P$ and $M_P$ stand for the pseudoscalar meson pole ($\pi$, K) parameters, and this FESR is suitable for FOPT. In the case of CIPT the FESR, Eq.(11), becomes
\begin{eqnarray}
- \frac{1}{2\pi i}
\oint_{C(|s_0|)}
&ds& \;\psi_{5}^{'' QCD}(s)\,[F(s) - F(s_0)] = 2\; f_P^2 \; M_P^4\; \Delta_5(M_P^2)  \nonumber \\ [.3cm]
&+&
\frac{1}{\pi} \; \int_{s_{th}}^{s_0}
ds \; Im \;\psi_{5}(s)|_{RES}\;\Delta_5(s) \, , 
\end{eqnarray}
where $F(s)$ is defined in Eq.(10).
It should be clear that the distortion introduced by the integration kernel $\Delta_5(s)$, Eq.(17), affects all three terms of the FESR as Cauchy's theorem remains valid as long as the kernel is an analytic function. Hence, the suppression achieved in the hadronic resonance contribution is compensated by corresponding changes in the pseudoscalar meson pole  and in the QCD contributions. Since these two terms are reasonably well known, this is a welcome feature.
\section{RESULTS}
The light-quark pseudoscalar correlator in PQCD is known up to fifth-loop order \cite{BAIKOV2}, with the strong coupling being determined from data on $\tau$-decays \cite{ALEPH}-\cite{PICH}, $\alpha_s(M_\tau^2) = 0.344 \pm 0.009$ which corresponds to a QCD scale in the $\overline{MS}$ scheme of $\Lambda = 365 - 397 \;\mbox{MeV}$. The handling of logarithmic quark mass singularities in this correlator requires some care, as explained in \cite{JAMIN}-\cite{CHET}.
The leading non-perturbative contributions of dimension $d=4$  are also known, together with higher order quark mass corrections. Complete expressions of all these contributions to the FESR for both FOPT and CIPT may be found in \cite{DNRS1}-\cite{DNRS2}. A posteriori, it is found that quark mass terms of order $\cal{O}$$(m^4)$ and higher, as well as vacuum condensates of dimension $d\geq 6$ are negligible on account of the integration kernel, Eq.(17). The hadronic resonance spectral function has been modeled by two Breit-Wigner forms normalized at threshold according to chiral perturbation theory \cite{CAD0}-\cite{PRADES}. The quark masses were determined with and without this contribution in order to gauge its impact on the final result. It turns out that allowing for a $\pm\,30\%$ uncertainty in the hadronic resonance contribution impacts on the quark masses at the level of less than 1\%. Assuming the unknown six-loop PQCD contribution to be equal to the known five-loop term also has an impact at the 1\% level. The presence of the integration kernel, Eq.(17), thus enhances considerably the importance of the better known contributions to the FESR, i.e. PQCD and the pseudoscalar meson pole. The major sources of uncertainty are then the values of the strong coupling and of the radius of the integration circle $s_0$. Results for the strange quark mass in FOPT are shown in Fig.6. The width of the stability region is typical of FESR applications. However, in CIPT as shown in Fig.7 this region is remarkably wide, with $m_s(2 \;\mbox{GeV})$ being exceptionally stable. This is also the case for the up- and down-quark masses in CIPT as shown in Figs. 8 and 9. While there is no clear cut criterion to establish which integration technique is best, results from this application clearly favour CIPT in terms of the stability region. Taking into account all possible sources of uncertainty, and following a very conservative approach of adding them up, rather than combining them in quadrature, the final results (in the $\overline{MS}$ scheme) are
\begin{eqnarray}
m_u (2 \;\mbox{GeV}) = 2.9 \; \pm \; 0.2 \; \mbox{MeV} \;, \nonumber \\ [.2cm]
m_d (2 \;\mbox{GeV}) = 5.3 \; \pm \; 0.4 \; \mbox{MeV} \;, \nonumber \\ [.2cm]
m_{ud} \equiv \frac{m_u + m_d}{2} = 4.1 \; \pm \; 0.2 \; \mbox{MeV}\;, \nonumber\\ [.1cm]
m_s (2 \;\mbox{GeV}) = 102 \; \pm \; 8\; \mbox{MeV} \;. \nonumber 
\end{eqnarray}
This is at present the most accurate determination of the light quark masses using QCD sum rules, with the above results in very good agreement with the most recent lattice QCD values \cite{LAT1}-\cite{LAT6}. With the systematic uncertainties from the hadronic resonance sector under very good control, a future reduction of the errors in the quark masses will mostly rely  on more accurate determinations of the strong coupling.

\end{document}